\documentstyle[12pt]{article}

\begin{document}
\begin{center}
{\Large \bf Doped  Manganites Beyond Conventional Double-Exchange Model}

\vspace{0.5cm}

{\bf A.S. Moskvin, I.L. Avvakumov}

\vspace{0.5cm}

{\bf Department of Theoretical Physics, Ural State University,

Ekaterinburg, 620083, Russia}

\vspace{0.5cm}




{\Large  Abstract}
\end{center}

The problem of adequate electronic model for doped manganites like
$La_{1-x}Sr_{x}MnO_3$ remains controversial. There are many thermodynamic and
local microscopic quantities  that cannot be explained by the conventional
double-exchange model with dominantly $Mn3d$ location of doped holes. In
such a situation we argue a necessity to discuss all possible candidate states
with different valent structure of  manganese and oxygen atoms, as well as
different valent states of  octahedral $MnO_{6}$ centers.

In  frames  of rather conventional quantum-chemical approach, crystal field and the ligand
field model we address different types of $MnO_6$ centers, different types of
$d-d$, and charge  transfer  transitions. We draw special attention to the
so-called charge transfer states related to strong intra-center charge
fluctuations. As we conjecture, namely these  could become active valent states
for doped manganites. We discuss some electric and magnetic properties of the
electron $MnO_{6}^{10-}$, and hole $MnO_{6}^{8-}$ centers with unconventional
 ${}^{6}A_{1g}-{}^{6}T_{1u}$ and ${}^{4}A_{2g}-{}^{4}T_{2u}$
 valent manifolds, respectively.

 We propose two idealized theoretical models for hole system in doped manganites.
The first one implies an overall oxygen localization for the doped holes occupying
the non-bonding $O2p$ orbitals. The
second assumes  a doping induced formation of the electron-hole Bose liquid, or
a system of the electron $MnO_{6}^{10-}$, and hole $MnO_{6}^{8-}$ centers. In a
sense, this scenario resembles a well-known disproportionation reaction. In
both cases one might expect non-trivial magnetic behavior with strong
ferromagnetic fluctuations due to anomalously strong ferromagnetic coupling of
non-bonding $O2p$ holes with $Mn3d$ electrons.



\section{Introduction}
The discovery of the high-$T_c$ superconductivity in doped cuprates and
colossal magnetoresistance in doped manganites like $La_{1-x}Sr_{x}MnO_3$ have
generated a flurry of ideas, models and scenarios
 of these puzzling phenomena, many of which are being developed up to date, although the situation
 remains controversial in both cases. Observation of various unconventional properties in doped
 cuprates, manganites, bismuthates with  high-$T_c$'s, nickellates and many other oxides shows
  that we deal with a manifestation of novel strongly correlated states with a
  "metal-dielectric"  duality, strong coupling of different (charge, spin, orbital,
  structural) degrees of freedom and non-Landau behaviour of quasiparticle. Moreover,
  for such {\it strongly correlated oxides} one observes many features typical for boson
  or boson-fermion systems. Finally, in all cases we deal with a specific role of nonisovalent
  substitution which appears to be a trigger mechanism to form a novel phase.


Current approach to manganites implies an applicability of conventional band
model with some modifications resulting in what one calls the "double exchange
plus Jahn-Teller" (DE+JT)  model \cite{DE+JT}. The model is a generalization of
the well-known double exchange model   \cite{dblex,dblex1}
  and implies a predominantly $Mn3d$ character of valent states both for parent
  and hole-doped system. Indeed, according to the conventional band theory
  calculations \cite{Millis,Solovyev} the conduction band of $LaMnO_3$ is derived mainly from
  $Mn e_g$ symmetry $d$-orbitals and is well separated from other bands. So, the low
 energy ($\hbar \omega \leq 4$ eV ) physics of these materials is believed to be governed by
  $Mn\,e_g$ electrons, which are coupled by a strong Hund's coupling, $J_H$ to $Mn\,t_{2g}$
  symmetry "core spins" and also interact which each other and with lattice distortions
   \cite{Millis}. This model has been developed, in particular, with aim to relevantly describe
    the low frequency conductivity data \cite{Millis}. An appropriate effective Hamiltonian
    could be represented as a sum of several terms
\begin{equation}
\hat H = \hat H_{KE} + \hat H_{\mu} +\hat H_{JT} +\hat H_{Hund} +\hat H_C\, ,
\end{equation}
\begin{equation}
\hat H_{KE}+\hat H_{\mu}= \sum _{i,j,a,b,\nu}t_{ij}^{ab}{\hat
d}^{\dagger}_{ia\nu}{\hat d}_{jb\nu}+(h.c.) +\mu \sum _{i,a\nu}{\hat
d}^{\dagger}_{ia\nu}{\hat d}_{ia\nu},
\end{equation}
\begin{equation}
\hat H_{JT}=-\lambda \sum _{i\nu}{\hat d}^{\dagger}_{i\nu}({\hat \sigma}_{x}
Q_{2}+{\hat \sigma}_{z} Q_{3}){\hat d}_{i\nu},
\end{equation}
where $\lambda$ is a vibronic constant, ${\hat d}^{\dagger}_{i\nu}({\hat
d}_{i\nu})$ is a two-component ($a,b$) creation (annihilation) operator,
$\sigma _{x,z}$ Pauli matrices, $Q_{2,3}$ are oxygen displacement modes for
octahedral $MnO_6$ centers. The Hund coupling is written as follows
\begin{equation}
\hat H_{Hund}= \sum _{i,a} J_{H}({\hat {\vec s}}_{i}\cdot {\hat {\vec S}}_{i}),
\end{equation}
where ${\hat {\vec S}}_{i}$ represents the $t_{2g}$ core spin and ${\hat {\vec
s}}_{i}$ does the $e_g$ one; ${\hat H}_C$ describes the Coulomb correlation
interaction.

Unfortunately, this approach is likely to preserve all the shortcomings of the conventional
band models and cannot provide the correct description of the low energy
physics for doped manganites. Indeed, the energy spectrum  of $MnO_6$ center
with one and two $e_g$ holes has nothing to do with that
 of  $Mn^{3+}$ and $Mn^{2+}$, respectively. The authors \cite{Millis} did not solve
  the Jahn-Teller problem, and $\hat H_{JT}$ virtually appears to be simple
  low-symmetry crystalline field term. They
   neglect a rather strong exchange between $Mn^{4+}$ spin cores. In addition, the
strong intra-atomic Hund exchange coupling cannot be correctly described by the
mean-field approximation, that results in some doubts about calculated
temperature dependences. Traditional model approach reduces the role of the
oxygen ions to the indirect effects of the $Mn3d-O2p$ covalent bonding
resulting in crystal field effects and renormalization of the $Mne_g$ and
$Mnt_{2g}$ orbitals.

  In our opinion, conventional approach is questionable both
  in what concerns the electronic structure of valent states and electronic homogeneity
  in doped systems. There are many thermodynamic and local
  microscopic quantities  that cannot be explained by the conventional
   double-exchange model.

It should be noted that unconventional behavior for the most part of strongly
correlated oxides is generated by a nonisovalent substitution accompanied by
strong fluctuations of the electric potential and considerable charge
fluctuations. It seems highly probable that this substitution results in a
stabilization of fluctuations of a novel strongly polarizable metallic-like phase that provides the
most effective screening of charge inhomogeneity. Owing to the features of the
pinning potential, the novel phase could hardly resemble  the conventional
phase states of the  pure homogeneous  systems. This could form a large
non-adiabatic polaron or topological defect texture somewhat like a vortex. The
topological order   endows the doped manganite with tremendous amount of
robustness to various unavoidable "real-life" material complications, such as
impurities and other coexisting broken symmetries. Such a situation in
manganites implies a necessity to examine different valent states of manganese
and oxygen atoms, as well as different valent states of a slightly distorted
octahedral $MnO_{6}$ center being the basic unit for crystalline and electron
structure.

Specific property both of $Mn3d$ and $O2p$ holes in manganites is in their
strongly correlated nature, so  the conventional ($metallic-like$) text-book
band models like LDA or even its recent modification LDA+U cannot provide a
relevant description of electronic structure and
 energy spectrum. This situation implies a making use of $dielectric-like$ quantum-chemical
 approach as the most relevant basis for realistic description both of local intra-center and nonlocal
 intercenter correlations.
Natural candidate for an effective center in manganites is slightly distorted
octahedral $MnO_{6}$ center  being the basic unit for crystalline and
electronic structure.

In the doped manganites we have to deal with   strong fluctuations of  the
crystal fields. Moreover, in these     systems
    one might expect strong fluctuations, up to the sign reversal, of the charge
    transfer energy $\Delta _{pd}$, $e.g.$ due to the substitution driven fluctuations of
     Madelung potential. Such fluctuations result in a crucial modification of conventional
electron structure and energy spectrum for the cation-anion centers with
possible stabilization of the charge transfer (CT) states, or the oxygen
location of holes. The Figure 1 illustrates such a situation for the hole
$MnO_{6}^{8-}$ center. The $\Delta _{pd}$ parameter in the mixed
    valence systems appears to be the variable order parameter, in contrast to conventional
    oxides (fluorides,...), where it is quenched at certain positive value. It seems, many
    doped oxides such as cuprates, manganites represent systems with the self-organized fluctuations
     of this parameter. Many conventional approaches such as band
   model,
  crystal field theory and ligand field theory should be significantly modified
   to describe
  these systems with mixed, or strictly speaking, indefinite valence, and the strong
   fluctuations
  of electron potential comparable with its mean value. 

The unconventional oxygen nature of the doped holes manifests itself in many
experimental observations. In this connection one should mention optical data
\cite{Jung} concerning the doping induced red shift of the anion-cation CT
transitions,
 "arrested" $Mn^{3+}$-valence response in the  $Mn$ $K\beta $ emission spectra of
 $La_{1-x}Sr_{x}MnO_3$ to the doping in the $x<0.3$ range founded in Ref.
 \cite{Tyson}, anomalous magnitude of the effective magnetic moment per manganese
 ion that exceeds any thinkable theoretical value \cite{Korolyov,Mukhin}.


%


In addition, one should note that traditional approach to the doped manganites
with their colossal magnetoresistance as to conventional homogeneous media are
incomplete and misleading. As for other strongly correlated oxides a phase
separation and tendency to form different inhomogeneous structures seems to be
a generic property of manganites \cite{Nagaev,Moskvin,Khomskii,Kugel}.
Nonetheless, a most part of the existing scenarios for the  doped manganites
corresponds to the usual effective-medium models and does not take account of
their complicated inhomogeneous and, probably, phase-separated structure.

In present paper we discuss several candidate phase states of doped manganites
with different valent structure of  manganese and oxygen atoms, as well as
different valent states of  octahedral $MnO_{6}$ centers.


In Sec.II in the framework of the model of localized centers, crystal field and
the ligand field model we address different types of $MnO_6$ centers, different
types of $d-d$, and charge  transfer  transitions. We draw special attention to
the so-called charge transfer states related to strong intra-center charge
fluctuations. As we conjecture, namely these  could become active valent states
for doped manganites. Here we discuss some electric and magnetic properties of
$MnO_6$ centers with unconventional valent manifold.

In Sec. III after  short discussion of a conventional double-exchange model we
present a novel model of oxygen hole ferromagnetic metal.

In Sec. IV we  address a novel mechanism for the insulator-metal transition
that is well known for such standard semiconductors with filled bands as $Ge$
and $Si$. Under certain conditions these reveal a nucleation of metallic
electron-hole (EH) liquid  resulting from  exciton decay \cite{Rice}. We
speculate, that somewhat like this could take place in doped manganites where
the electron-hole system should be equivalent to a Bose-liquid.  In this
section we briefly touch upon an issue of analogy with chemical
disproportionation reaction, and experimental observation of different $Mn$
valence states in manganites.

\section{Electron structure of manganese ions and manganese-oxygen
octahedral centers in manganites}
 Five $Mn3d$ and eighteen $O2p$ atomic
orbitals in octahedral $MnO_6$ complex form both hybrid $Mn3dO2p$  bonding and
antibonding $e_g$ and $t_{2g}$ molecular orbitals (MO), and non-bonding
$a_{1g}(\sigma)$, $t_{1g}(\pi)$, $t_{1u}(\sigma)$, $t_{1u}(\pi)$, $t_{2u}(\pi)$
ones \cite{Ber}. {\it Conventional} electronic structure of octahedral $MnO_6$
complexes is associated with configuration of the completely filled $O2p$
shells and partly filled $Mn3d$ shells. {\it Unconventional} electronic
configuration of octahedral $MnO_6$ complexes is associated with a {\it charge
transfer (CT) state} with  one hole in $O2p$ shells. The excited CT
configuration $\; \tilde \gamma _{2p}^1 \, 3d^{n+1} \;$ arises from the
transition of an electron from the MO predominantly anionic in nature (the hole
$\,\tilde\gamma_{2p} \; $ in the core of the anionic MO being hereby produced),
into an empty $\,3d\,$-type MO ($t_{2g}\,$ or $\,e_g$). The transition between
the ground configuration and the excited one can be presented as the
$\,\gamma_{2p}\,\rightarrow\,3d\,(t_{2g}\,,\>e_g)\quad CT\,$ transition. The CT
configuration consists of two partly filled subshells, the ligand $\; \gamma
_{2p}\,$-, and the cation $\,3d\,(t_{2g}^{n_{1}}e_{g}^{n_{2}}$ shell,
respectively. It should be emphasized that  the oxygen hole having occupied the
{\it non-bonding} $\; \gamma _{2p}\,$ orbital interact {\it ferromagnetically}
with $\,3d\,(t_{2g}^{n_{1}}e_{g}^{n_{2}})$ shell. This rather strong (up to
$\sim 0.1$ eV) ferromagnetic coupling results in Hund rule for the CT
configurations, and provides the high-spin ground states.

\subsection{ $Mn^{3+}$ ion and $MnO_{6}^{9-}$ center }
\subsubsection{Conventional electronic structure}
 The typical high-spin ground state configuration and crystalline
term for $Mn^{3+}$ in octahedral crystal field or for the octahedral
$MnO_{6}^{9-}$ center is $t_{2g}^{3}e_{g}^1$ and ${}^{5}E_{g}$, respectively.
Namely this orbital doublet results in a vibronic coupling and Jahn-Teller (JT)
effect for the $MnO_{6}^{9-}$ centers, and cooperative JT ordering in
$LaMnO_3$. In the framework of crystal field model the ${}^{5}E_{g}$ term
originates from the  $(3d^{4}\,{}^{5}D)$ term of  free $Mn^{3+}$ ion.

Among the low-energy crystal field d-d transitions for the high-spin $Mn^{3+}$
ions one should note a single spin-allowed and parity-forbidden
${}^{5}E_{g}-{}^{5}T_{2g}$ transition at energy varying from about $2$ eV to
$2.5$ eV depending on the crystalline matrix. So, this is $2.5$ eV for the
$Mn^{3+}$ impurity in perovskite $YAlO_3$ \cite{Gen}.
 The transition is magneto-optically active, and could manifest itself in the Faraday and
  Kerr effects. A detection of the spin- and parity-forbidden ${}^{5}E_{g}-{}^{3}T_{1g}$
  transition with, probably, lower energy could be rather important in  the $Mn^{3+}$
  assignment. Some authors   consider the orbital and parity-forbidden transition
  ${}^{5}E_{g}^{'}-{}^{5}E_{g}^{''}$ between two components of the ${}^{5}E_{g}$ doublet
   splitted by the low-symmetry crystalline field to be an origin of a rather strong absorption
    observed in  manganites near $1.5$ eV.
Such a rather big  magnitude of the splitting for the ${}^{5}E_{g}$ doublet
agrees well with observed magnitudes  of the low-symmetry distortions for  the
$MnO_{6}^{9-}$ octahedra  of the order of $15\%$ \cite{Zhou}.

\subsubsection{Intra- and inter-center charge transfer transitions for the
$MnO_{6}^{9-}$ centers. Small charge transfer excitons}

A set of the intensive and broad absorption bands in parent manganites is
related to the anion-cation $O2p-Mn3d$ charge transfer. In the framework of the
$MnO_{6}$ center model this elementary CT process generates both intra- and
inter-center CT transitions for the $MnO_{6}^{9-}$ centers. The intra-center CT
transitions, in  turn, could be subdivided \cite{Davydov} to {\it localized
electronic excitations} and {\it Frenkel excitons},
 respectively. The inter-center CT transitions form a set of {\it small CT  excitons},
  which  in terms of chemical notions represent somewhat like the {\it disproportionation}
  quanta
\begin{equation}
  MnO_6^{9-}+MnO_6^{9-} \rightarrow MnO_6^{10-}+MnO_6^{8-}
  \label{dispro}
\end{equation}
resulting in a formation of the bounded electron $MnO_6^{10-}$  ($e$-) and hole
$MnO_6^{8-}$
  ($h$-) small radius centers.  A minimal energy of such an exciton or the disproportionation
   threshold turns out to be rather small ($\leq 5.0$ eV) as compared with appropriate purely
   ionic quantity which value equals to the electrostatic correlation energy $U_{dd}\approx
   10$ eV.

{\bf Intracenter charge transfer transitions for the $MnO_{6}^{9-}$ centers.}

Conventional classification scheme of the charge-transfer transitions in the
octahedral $MnO_{6}^{9-}$ centers (intra-center CT transitions) incorporates
the electro-dipole allowed
 $(t_{2g}^{3}e_{g}^1){}^{5}E_{g}-$ $(t_{2g}^{4}e_{g}^{1}\underline{\gamma _{u}}){}^{5}T_{1,2u},
 $ $ (t_{2g}^{3}e_{g}^{2}\underline{\gamma _{u}}){}^{5}T_{1,2u}$ transitions from the odd oxygen
 $\gamma =5t_{1u},6t_{1u},1t_{2u}$ orbitals to the even manganese $3dt_{2g}$ and $3de_g$
 orbitals, respectively. Most probably, the minimal energy separation is related to the CT
 transition  $1t_{2u}-e_g$ with formation of the excited $Mn^{2+}$-like $t_{2g}^{3}e_{g}^{2}$
  configuration. As for manganite $LaMnO_3$, the minimal energy of the allowed CT transitions
   is estimated to be  $\Delta _{CT} \geq 4.0$ eV. In addition to the allowed CT transitions
    one should note a number of the parity-forbidden CT transitions, mainly the $1t_{1g}-t_{2g},
    e_{g}$ ones with the lowest energy.

\subsection{The hole $MnO_{6}^{8-}$ center in doped manganites:
conventional text-book  model}

 The hole
centers in manganites may be created as a result of the photo- or chemical
doping  as in  $La^{3+}_{1-x}Sr^{2+}_{x}MnO_3$. In both cases the hole
$MnO_{6}^{8-}$ center can be
 chosen as a relevant model entity to describe a real distribution of the hole density.

Conventional model approach to the
hole $MnO_{6}^{8-}$ center implies $\Delta _{pd}>0$ and the
 ground state with predominantly $Mn3d$ hole location (see Fig.1).
 The typical high-spin ground state configuration
and crystalline term for $Mn^{4+}$ in octahedral crystal field or for the
octahedral $MnO_{6}^{8-}$ center is half-filled configuration $t_{2g}^{3}$ and
the orbital singlet  term ${}^{4}A_{2g}$, respectively.
 In the framework of crystal field model the ${}^{4}A_{2g}$ term originates from the
  $(3d^{3}{}^{4}F)$ term of  free $Mn^{4+}$ ion.
Among low-energy crystal field d-d transitions for $Mn^{4+}$ ions one should
note a rather
 intense (oscillator strength $\sim 10^{-3}$) parity-forbidden ${}^{4}A_{2g}-{}^{4}T_{2g}$
  transition at energy varying from about $2$ eV to $2.6$ eV depending on the crystalline
  matrix. The transition is magneto-optically active, and could manifest itself in the Faraday
   and Kerr effects. The very weak spin- and parity-forbidden
   ${}^{4}A_{2g}-{}^{2}E_{g},{}^{2}T_{1g}$ transitions can be observed at  lower energy
   $1.5\div 2$ eV. The relatively weak orbital and parity-forbidden ${}^{4}A_{2g}-{}^{4}T_{1g}$
   transition has energy about $3$ eV.

The classification scheme for the CT transitions in octahedral $MnO_{6}^{8-}$
centers is similar
 to that of $MnO_{6}^{9-}$ centers, described above. As for manganite
 $LaMnO_3$\cite{Paulusz},
  the minimal energy of the allowed CT transitions is estimated to be
  $\Delta _{CT} \sim 4.0$ eV.
 The luminescence measurements in \cite{MnO}  give the  energies  $2.6$ eV and $3.1$ eV for
  ${}^{4}A_{2g}-{}^{4}T_{2g}$ and ${}^{4}A_{2g}-{}^{4}T_{1g}$ transitions, respectively.

Very important information about the octahedral $Mn^{4+}$ spectra with likely
no problems of mixed Mn valence  can be deduced from the magneto-optical
studies of manganese pyrochlores $A_{2}^{3+}Mn_{2}^{4+}O_7$ ($A=In,\,
Tl,\,Lu,\,Yb$) \cite{Balykina}. The authors observed two rather narrow
magneto-optical features in the range $2.6-3.1$ eV, related to the $d-d$
 transitions ${}^{4}A_{2g}-{}^{4}T_{2g}$ and ${}^{4}A_{2g}-{}^{4}T_{1g}$, respectively,
 and two rather wide features near $3.7-3.9$ eV and $\geq 4.3$ eV, associated with CT
 transitions. This assignment agrees with data by Paulusz and Burrus \cite{Paulusz},
  and Jung {\it et al.} for $CaMn^{4+}O_3$ \cite{Jung}.

\subsection{The electron $MnO_{6}^{10-}$ center in doped manganites:
conventional scheme}
The typical high-spin ground state configuration and
crystalline term for $Mn^{2+}$ in octahedral crystal field or for the
octahedral $MnO_{6}^{10-}$ center is a half-filled configuration
$t_{2g}^{3}e_{g}^2$ and  the orbital singlet  term ${}^{6}A_{1g}$,
respectively.
  In the framework of crystal field model this term originates from the $(3d^{5}{}^{6}S)$ term
  of  free $Mn^{2+}$ ion. It should be noted that for the $3d^{5}$ configuration the
  ${}^{6}A_{1g}$ term is the only spin-sextet one, so all the $d-d$ transitions from the
   ground state are spin- and parity-forbidden, therefore these are extremely weak (oscillator
    strength $\sim 10^{-7}$).
Among low-energy crystal field $d-d$ transitions for $Mn^{2+}$ ions one should
note a sequence
 of ${}^{6}A_{1g}-{}^{4}T_{1g},{}^{4}T_{2g},({}^{4}A_{1g},{}^{4}E_{g})$ transitions to be
  observed in the spectral range   $2\div 3$ eV depending on the crystalline matrix.
 In manganese oxide $Mn^{2+}O$ these transitions are observed at $\approx 2.0$ eV and
 $\approx 2.95$ eV for ${}^{4}T_{1g}$ and ${}^{4}A_{1g},{}^{4}E_{g}$, respectively \cite{MnO}.
 All the  transitions are magneto-optically active, and could manifest themselves in the Faraday and
  Kerr effects.

The classification scheme for the CT transitions in octahedral $MnO_{6}^{10-}$
centers is similar to that of $MnO_{6}^{9-}$ centers.

 The many-electron state of the CT
configuration is $\; \tilde \gamma _{2p}^1 \, : \, t_{2g}^4 \, e_g^2 (^5T_2)\,
: \, ^{2S+1} \Gamma \;$ or $\; \tilde \gamma_{2p}^1\,:\,t_{2g}^3 e_g^3
\,(^5E)\,:\,^{2S+1} \Gamma \;$ (due to the transfer $\; \gamma _{2p}
\rightarrow \, t_{2g} \;$ and $\; \gamma _{2p} \, \rightarrow \, e_g\,$,
respectively).  The symmetry selection rules for electro-dipole transitions
 permit only the following values of $\; \gamma \>$ \cite{Ber}: $\; \gamma
\, = \, t_{1u}\;$ (two MO usually denoted as $\; t_{1u} (\pi)\;$ and $\; t_{1u}
(\sigma)\,$) and $\; \gamma \,=\, t_{2u}\;$ (single $\;MO$). In accordance with
thinkable variants of the combination of "intermediate" quantum numbers of the
oxygen and manganese subsystems, six "one-electron" CT transitions (the lowest
in energy) $\; t_{2u}, \; t_{1u}(\pi), \; t_{1u}(\sigma) \> \rightarrow \>
t_{2g}\;$  and $\; t_{2u}, \; t_{1u}(\pi), \; t_{1u}(\sigma)\> \rightarrow \>
e_g \;$ give rise to a lot of "many-electron" $\;CT\;$ transitions, among which
only six transitions of the $\; ^6A_{1g}\,- \, ^6T_{1u} \>$ type are
electric-dipole allowed, and others are forbidden either by the quasimomentum
selection rule (transitions of the $\; ^6A_{1g} \, - \, ^6 \Gamma _u \>$-type,
$\; \Gamma \, \not= \, T_1 $) or due to spin (transitions of the $\; ^6A_{1g}
\, - \, ^4 \Gamma _u \,$-type; if $\; \Gamma \, \not= \, T_1 \>$, they are
doubly forbidden).

A rather detailed description of the optical and
 magneto-optical manifestation of similar transitions in isoelectronic octahedral centers
 $FeO_{6}^{9-}$ for some ferrites  is made in Ref.\cite{Zenkov}. Due to a lower valence of cation $Mn^{2+}$
  the CT bands in $MnO_{6}^{10-}$ can be shifted to bigger energies with the CT gap $\sim 4.0 $
  eV.

\subsection{Unconventional $MnO_{6}$ centers.
The   valence crossover and pseudo-Jahn-Teller centers}

\subsubsection{Isolated $MnO_6$ centers}

 A significant magnitude of the CT energy $\Delta _{pd}\geq 2$ eV is
a typical situation for the conventional transition metal oxides. However, this
situation can be altered, in particular, for the intermediate (mixed) valence
systems with strong fluctuations of
 electron potential, where an effective  CT energy $\Delta _{CT}$ could reach very small
 values, and even to be negative one. Such a case implies  close energies, or near-degeneracy,
  for two states with a hole localized predominantly on the $3d$ cation and the one localized
  on either purely oxygen orbital, respectively. This quasi-degeneracy or
  $Mn^{3+}-Mn^{4+}$  valence resonance (see Fig.1) could result in the formation of the
  hole $MnO_{6}^{8-}$ pseudo-Jahn-Teller center, or small PJT polaron. The PJT polarons will
   manifest themselves in various physical effects. 
     
     One of the most probable  bare electronic states for the hole PJT center  $MnO_{6}^{8-}$
are $(t_{2g}^{3}){}^{4}A_{2g}$ and 
$(t_{2g}^{4}e_{g}^{2}\underline{t}_{2u}){}^{4}T_{2u}$ one.
Indeed, these even and odd states are most strongly coupled due to dipole-charge interaction
with dipole-active displacement modes for $MnO_6$ octahedra that could
result in a strong
PJT effect with a multi-well adiabatic potential. Formally, such a center looks like
 a so-called charge transfer vibronic exciton (CTVE)\cite{Vikhnin}, or charge transfer
 excitation $(t_{2g}^{3}){}^{4}A_{2g}-(t_{2g}^{4}e_{g}^{2}\underline{t}_{2u}){}^{4}T_{2u}$
 strongly stabilized by a local  dipole-active structural distortion. The important
 feature of such a PJT center, or CTVE, is the oxygen-manganese charge transfer
 which is self-consistently coupled with a local   structural distortion. In other
 words, such a center could be described as a system of two coupled dipole moments
 generated by charge transfer and ionic displacements, respectively.

      One should note that $(t_{2g}^{3}){}^{4}A_{2g}$ and 
$(t_{2g}^{4}e_{g}^{2}\underline{t}_{2u}){}^{4}T_{2u}$ states correspond to different $3d$ configurations: $3d^3$ and
       $3d^4$, familiar to $Mn^{4+}$ and $Mn^{3+}$ ions, respectively. This results in an
       unconventional $d-d$ spectrum for the hole PJT center $MnO_{6}^{8-}$, it will represent
        a superposition of $d-d$ transitions for $Mn^{4+}$ and $Mn^{3+}$ ions with relative
        intensity specified by the relative weights of bare electronic states in vibronic
         ground state. A $Mn^{3+}$-like $d-d$ spectrum for the hole PJT center $MnO_{6}^{8-}$
          has to be distinguished from the  $d-d$ spectrum for the $Mn^{3+}$ in the proper
$MnO_{6}^{9-}$ center. An occurrence of additional oxygen hole in the
            former center has to result in a screening effect with a lowering of the crystal field splitting
parameter $10Dq$.

The $\Delta _{pd}$ sign crossover results in a peculiar interchange of valent
$Mn3d$ holes to valent $O2p$ holes that generates a number of novel puzzling
phenomena such as   oxygen anionic magnetism, oxygen anionic metal,
oxygen-oxygen CT transitions.

In optics  the PJT center manifests itself
   in mid-infrared (MIR) absorption related to transitions between vibronic states for different
   sheets of adiabatic potential \cite{Ber,Moskvin1993,Moskvin1994}. The MIR
   absorption band will be especially intensive
when the allowed electro-dipole transition  between the  bare electronic states
is possible.

{\bf Unconventional $MnO_{6}^{10-}$ center:$Mn^{2+}-Mn^{1+}$ ionic-covalent
resonance.}

 Similarly to the hole $MnO_{6}^{8-}$ center we can
expect formation of the electron $MnO_{6}^{10-}$ PJT center given certain
conditions. This is a center with nominal $Mn$ valence
 resonating between $Mn^{2+}$ and $Mn^{1+}$.
The mid-infrared region transitions and $Mn^{1+}$-like d-d transitions
represent a specific optical  manifestation of the electron PJT center. The
$Mn^{1+}$ $3d^6$ configuration has the high-spin ground state Hund term
${}^{5}D$, or crystalline term ${}^{5}T_{2g}$  for  octahedral crystal field
with a rather intense both
 optically and magneto-optically spin-allowed ${}^{5}T_{2g}-{}^{5}E_{g}$ transition. However,
  the most typical situation for  octahedral centers with the $3d^6$ configuration corresponds
  to the competition of the high- and low-spin ground states with filled $t_{2g}$ shell
  $(t_{2g}^{6}){}^{1}A_{1g}$. Then one observes a rather intense spin-allowed
  ${}^{1}A_{1g}-{}^{1}T_{1g}$ transition with energy $2\div 3$ eV and
  ${}^{1}A_{1g}-{}^{1}T_{2g}$ at a slightly ($\sim 0.5$ eV) higher energy.
  In addition, there   exist numerous weak  spin- and parity-forbidden transitions.

\subsubsection{The $MnO_6$ centers in a polarizable lattice}

Above we implied an isolated $MnO_6$ center with varying $\Delta _{pd}$ parameter. However, for $MnO_6$ center in a polarizable medium, such as perovskite lattice we should account for additional polarization effects. 
The perovskite structure is typical for ferroelectrics like $BaTiO_3$, and 
incipient ferroelectrics or quantum paraelectrics such as $SrTiO_3$. 
Soft and highly polarizable perovskite lattice together with valence instability
 or the ability of $Mn$ to exhibit intermediate valence states involving the 
 oxidation states $Mn^{2+}/Mn^{4+}$ together with  $Mn^{3+}$ could result in 
 the specific instability of the bare orbitally nondegenerate ground states 
 like $^{4}A_{2g}$ in $Mn^{4+}$ or $^{6}A_{1g}$ in $Mn^{2+}$ with regard to the
  cation-anion charge transfer accompanied by strong electric dipole moment 
  fluctuations. Indeed, in strongly polarizable media the GS orbital singlets like $(t_{2g}^{3}){}^{4}A_{2g}$ in $Mn^{4+}$ or $(t_{2g}^{3}e_{g}^{2}){}^{6}A_{2g}$  in $Mn^{2+}$
could be strongly coupled with the charge transfer states like  $(t_{2g}^{4}e_{g}\underline{t}_{2u}){}^{4}T_{2u}$
 and $(t_{2g}^{4}e_{g}^{2}\underline{t}_{2u}){}^{6}T_{1u}$, respectively, 
 due to the dipole-dipole polarization interaction with a possible formation
 of a peculiar cloud of the electronic polarization of the lattice.
  The dipole moment related to the $A_{g}-T_{u}$ manifolds could attain a 
  rather big value: $d_{AT}\approx eR_{MnO}\lambda$, especially for the 
  enhanced covalency effects realized for rather low $\Delta _{pd}$ values.

Thus we come to the formation of the specific charge transfer dipole center in a polarizable
lattice with the
mixed ionic-covalent type of the chemical bonding (ionic-covalent resonance) 
and fluctuating reorienting dipole moment. This situation is the straightforward 
analog of above mentioned PJT effect due to vibronic coupling with dipole-active 
lattice displacements. Moreover, both effects supplement each other.
In addition, it should be noted that close energy separation of even $A_g$-like
and odd $T_u$-like states with large magnitude of appropriate dipole matrix element
implies a large electron polarizability of such a center, so, a tend to clusterization.
Anyway, one of the most important consequences of this dipole instability for the 
systems of $MnO_6$ centers implies the incorporation of the odd charge transfer
 states like $(t_{2g}^{4}e_{g}\underline{t}_{2u}){}^{4}T_{2u}$
 and $(t_{2g}^{4}e_{g}^{2}\underline{t}_{2u}){}^{6}T_{1u}$ to the active 
 valent electronic states on equal footing with the bare ground state 
 like $(t_{2g}^{3}){}^{4}A_{2g}$  or $(t_{2g}^{3}e_{g}^{2}){}^{6}A_{2g}$, 
 respectively.
 
 Naturally, that for a correct self-consistent description of systems with
the effective $\Delta _{pd}$ sign crossover, or ionic-covalent resonance,
one should both substantially extend the valent electron manifold and include 
the lattice degrees of freedom. Moreover, because of the particular importance of 
polarization effects, rigorously speaking, we should introduce these to 
a modified effective 
Hamiltonian  having abandoned an attempt to make use of the conventional scheme
with effective correlation parameters like $U_d$ and $U_p$.

\subsection{Electric and magnetic properties of the charge transfer
$T_{1u,2u}$ states and $A-T$ valent manifolds for $MnO_6$ centers.}

The singlet Hund terms $^{6}A_{1g}$ and $^{4}A_{2g}$ for the centers
$MnO_{6}^{10-}$ and $MnO_{6}^{8-}$ originate from the electronic configurations
$t_{2g}^{3}e_{g}^{2}$ and $t_{2g}^{3}$, respectively. The orbital triplets
$^{6}T_{1u}$ and $^{4}T_{2u}$ are typical for several configurations with the
ligand-metal charge transfer. For the  $MnO_{6}^{10-}$ center one has the CT
configurations $t_{2g}^{4}e_{g}^{2}\b{t}_{2u}$,
$t_{2g}^{4}e_{g}^{2}\b{t}_{1u}(\pi)$, $t_{2g}^{4}e_{g}^{2}\b{t}_{1u}(\sigma)$,
related to the transfer of the oxygen $t_{2u},t_{1u}(\pi),t_{1u}(\sigma)$
electron into the predominantly $3dt_{2g}$ state. Similarly, the electronic
configurations $t_{2g}^{3}e_{g}^{3}\b{t}_{2u}$,
$t_{2g}^{3}e_{g}^{3}\b{t}_{1u}(\pi)$, $t_{2g}^{3}e_{g}^{3}\b{t}_{1u}(\sigma)$
correspond to the transfer of the oxygen electron to the cation $e_{g}$ state.

Below we address some electric and magnetic properties of the charge transfer
$T_{1u,2u}$ states and $A-T$ valent manifolds for $MnO_6$ centers.

\subsubsection{Electric properties.}
First, one should note the significant difference between the electron density
distributions in the orbital singlet $^{6}A_{1g}$($^{4}A_{2g}$) and in the
orbital triplet $^{6}T_{1u}$($^{4}T_{2u}$), respectively.

 In the orbital triplet state
$^{6}T_{1u}$($^{4}T_{2u}$) it is possible to introduce the effective orbital
moment $\tilde L =1$, which has the very simple matrix
\begin{equation}
\langle i|\tilde L _{k}|j\rangle=-i\epsilon _{ikj}.
\end{equation}
on the basis of the $|x,y,z\rangle$-states.

 The effective quadrupole moment of the
$^{6}T_{1u}$($^{4}T_{2u}$) term has the form
\begin{equation}
\hat Q^{\Gamma}_{ij}=Q_{\Gamma}(\widetilde {L_{i}L_{j}}-\frac{1}{3}{\vec
L}^{2}\delta _{ij}),
\end{equation}
where $\Gamma =E,T_2$, $Q_{\Gamma}$ is the  quadrupole parameter, $\widetilde
{L_{i}L_{j}}=\frac{1}{2}(L_{i}L_{j}+L_{j}L_{i})$.

 The nonzero matrix elements of the effective dipole moment for the valent multiplet are written as follows:
\begin{equation}
\langle ^{6}A_{1g}(^{4}A_{2g})|d
_{k}|^{6}T_{1u}(^{4}T_{2u})j\rangle=d_{A-T}\delta _{kj}.
\end{equation}

\subsubsection{Magnetic properties}
{\bf Effective magnetic moment.}

One of the most important magnetic peculiarities for the charge transfer states like $(t_{2g}^{4}e_{g}\underline{t}_{2u}){}^{4}T_{2u}$
 and $(t_{2g}^{4}e_{g}^{2}\underline{t}_{2u}){}^{6}T_{1u}$ is related to a nonquenched orbital magnetic moment. In common, the effective magnetic moment of the $MnO_6$ center can be written as follows
\begin{equation}
{\vec M}=\beta _{e}(g_{S}{\vec S}+g_{L}{\vec L}),
\end{equation}
where $g_S \approx 2$ is the spin $g$-factor, $g_L$ is the effective orbital
$g$-factor, which is represented as the sum of the $Mn3d$- and
$O2p$-contributions
\begin{equation}
g_{L}=g_{L}^{3d}+g_{L}^{2p}.
\end{equation}
The values of the parameters $g_L$ for all six different charge transfer
configurations of the  $MnO_{6}^{10-}$ and $MnO_{6}^{8-}$ centers resulting in 
$^{6}T_{1u}$ terms were
calculated in \cite{Zenkov}. The main charge-transfer configuration
$t_{2g}^{4}e_{g}^{2}\b{t}_{2u}$ for the $MnO_{6}^{10-}$-center turns out to
have the negative effective orbital $g$-factor
\begin{equation}
g_{L}=g_{L}^{3d}+g_{L}^{2p}=-\frac{1}{2}-\frac{1}{4}.
\end{equation}
The effective spin-orbital interaction has the standard form
\begin{equation}
V_{SO}=\lambda {\vec L}{\vec S},
\end{equation}
where $\lambda = \lambda ^{3d}+\lambda ^{2p}$ is the effective spin-orbital
constant. The magnitude of these parameters for all  above addressed electronic
configurations of the  $MnO_{6}^{10-}$ and $MnO_{6}^{8-}$ centers were
calculated in \cite{Zenkov}. For the main charge transfer configuration
$t_{2g}^{4}e_{g}^{2}\b{t}_{2u}$ of the $MnO_{6}^{10-}$ center
\begin{equation}
\lambda =\frac{1}{20}\xi _{3d}+\frac{1}{10}\xi _{2p}>0 \,.
\end{equation}
This corresponds to the antiparallel (antiferromagnetic) orientation of the
spin and the orbital momenta in the ground state of the term
$(t_{2g}^{4}e_{g}^{2}\b{t}_{2u}){}^{6}T_{1u}$. With the negative value of the
orbital $g$-factor this corresponds to the parallel (ferromagnetic) orientation
of the spin magnetic moment and the orbital magnetic moment. The maximal value
of the total magnetic moment is
\begin{equation}
M({}^{6}T_{1u})=5\beta _{e}+0.75\beta _{e}=5.75\beta _{e}\, ,
\end{equation}
that significantly exceeds the pure spin value.

{\bf Magnetic anisotropy.}

The source of the single-ion orbital magnetic anisotropy for the $MnO_6$
centers is the low-symmetry crystal field, which can be described by the
effective Hamiltonian
\begin{equation}
\hat H_{LSCF}=\sum _{ij}K_{ij}(\widetilde {L_{i}L_{j}}-\frac{1}{3}{\vec
L}^{2}\delta _{ij}).
\end{equation}
Here $K_{ij}$ are the effective constants of the single-ion orbital magnetic
anisotropy. The magnitudes of these parameters can vary in a sufficiently wide
region, taking the values $0.01\div 0.1$ eV for the low-symmetry lattice
distortions being of the order of $0.01\div 0.1$.

Two-ion orbital anisotropy is determined by the quadrupole-quadrupole
interaction and also can take large values. Thus, the $MnO_6$ centers in the
 case of the model manganite can exhibit large magnetic anisotropy,
compared with that of  the rare-earth compounds.

\section{Unconventional model approaches to electron structure of doped manganites}

\subsection{The oxygen hole ferromagnetic metal}
One of the simplest unconventional model approaches to the hole doped
manganites like $La_{1-x}Sr_{x}MnO_3$
 implies purely oxygen location of the doped holes. This result in an
 unconventional system with two, $Mn3d$ and $O2p$, non-filled shells.
 An appropriate effective Hamiltonian  could be represented as a sum of several terms
\begin{equation}
\hat H = \hat H_{KE} + \hat H_{\mu} +\hat H_{LSCF} +\hat H_{exch} +\hat H_C +
\hat H_{imp}\, ,
\end{equation}
\begin{equation}
\hat H_{KE}+\hat H_{\mu}= \sum _{i,j,\alpha ,\beta ,\nu}t_{ij}^{\alpha
\beta}{\hat p}^{\dagger}_{i\alpha \nu}{\hat p}_{j\beta \nu}+(h.c.) -\mu \sum
_{i,\alpha \nu}{\hat p}^{\dagger}_{i\alpha \nu}{\hat p}_{i\alpha \nu},
\end{equation}
where $\alpha ,\beta $ labels different ($x,y,z$) $O2p$ orbitals,
$t_{ij}^{\alpha \beta}$ are $O2p-O2p$ transfer integrals, $\mu$ chemical
potential.
\begin{equation}
\hat H_{LSCF}=\sum _{i\alpha ,\beta \nu}b_{\alpha \beta}(i){\hat
p}^{\dagger}_{i\alpha \nu}{\hat p}_{i\beta \nu},
\end{equation}
where $b_{\alpha \beta}(i)$ are the low-symmetry crystalline field parameters.
The exchange coupling incorporates the $d-d$,  $p-p$ and the strongest $p-d$
contributions, and is written as follows
\begin{equation}
\hat H_{exch}= \sum _{i,j} [J_{ij}^{dd}({\hat {\vec S}}_{i}\cdot {\hat {\vec
S}}_{j})+J_{ij}^{pd}({\hat {\vec s}}_{i}\cdot {\hat {\vec
S}}_{j})+J_{ij}^{pp}({\hat {\vec s}}_{i}\cdot {\hat {\vec s}}_{j})],
\end{equation}
where ${\hat {\vec S}}_{i}$ represents the $Mn^{3+}$  spin and ${\hat {\vec
s}}_{i}$ does the $O2p$ hole one. The terms $H_C$ and $H_{imp}$ describe the
intra- and inter-site Coulomb $O2p-O2p$ interaction and electrostatic
interaction with $Sr^{2+}$ induced impurity potential. One should note that
despite a wide-spread opinion the correlation effects for oxygen holes could be
rather strong \cite{Hirsch}.
 These should provide a coexistence  of the two 
(manganese and oxygen) non-filled bands. 

This  Kondo lattice ($p-d$) model with ferromagnetic $p-d$ coupling immediately
explains many unconventional properties of the hole doped manganites. First of
all, at low hole content we deal with hole localization in impurity potential.
Then, given further hole doping a percolation threshold occurs accompanied by
insulator-anionic oxygen metal phase transition and ferromagnetic ordering both
in oxygen and $Mn$ sublattices, due to  a strong ferromagnetic Heisenberg $pd$
 exchange. However, it should be noted that ferromagnetic sign of $pd$ exchange
 is characteristic of non-bonding $p$ and $d$ orbitals.

The oxygen hole doping results in a strong spectral weight transfer from the
intense $O2p-Mn3d$  CT transition    bands to the $O2p$ band developed. The
$Mn^{3+}$ $d-d$ transitions will gradually shift to the low energies due to a
partial $O2p$-hole screening of the crystalline field. In a whole, optical data
evidence in favor of the oxygen hole scenario. In addition, one might point to
the two exciting  experimental facts. Firstly, an "arrested" $Mn^{3+}$-valence
response in the $Mn$ $K\beta $ emission spectra of $La_{1-x}Sr_{x}MnO_3$ to the
doping in the $x<0.3$ range founded in Ref. \cite{Tyson} is  consistent with
creation of predominantly oxygen holes. Secondly, this is observation of
anomalously large magnitude of saturated magnetic moments in ferromagnetic
state for different doped manganites \cite{Korolyov,Mukhin}.

Coexistence of two non-filled non-bonding bands is a rather unconventional
phenomenon for usual homogeneous systems. Anycase this implies a
quasi-degeneracy in energy and instabilities in the system.

In connection with the oxygen location of doped holes in manganites we should
note  the  probable manifestation of the specific correlations inherent in
almost filled electron shells  \cite{Hirsch}. Indeed,  the conventional
approach to the electron structure of the cation-anionic clusters like
$CuO_4$-centers implies allocation of a rather small number (one, two) of
valent holes against the non-degenerate structureless (rigid) cation-anionic
background with full neglect of appropriate correlational coupling "valent hole
$-$ many-electron background". However, such a coupling could play an important
role, in particular, for anions like oxygen with almost filled $2p$-shell,
resulting in many quantitative and qualitative effects, including the
$O2p$-hole localization. Firstly, this implies a nonrigid degenerate structure
for anionic $O2p^6$ background with internal  degrees of freedom. Valent
hole(s) moves around this nonrigid background. Namely such a situation occurs
in a generalized "shell-droplet" model for nuclei after Bohr and Mottelson
\cite{Nataf}. It seems, a rather similar picture one observes for clusters with
the Jahn-Teller effect when an electron moves around nonrigid nuclear
configuration. Partly, this resembles  familiar transition from the rigid-ion
model to the shell  model in lattice dynamics.

The simplest "single-doublet"  model  was proposed by Hirsch et al.
\cite{Hirsch}
 who assumed
an occurrence of the doublet structure  of isolated anionic $O2p^6$ background
in copper oxides
 and made use of the pseudo-spin formalism. Unfortunately, this very fruitful
  approach remained
 to be only interesting idea despite the ever growing interest in electronic
  structure and
 correlations in different oxides.
The detailed theoretical treatment of the unconventional anionic oxygen hole
model is needed to provide an appropriate description of the optical, magnetic,
and many other physical properties.

\subsection{The electron-hole Bose liquid ("disproportionation") scenario  in
manganites} Above we considered two opposite versions of charge fluctuations in
the hole doped manganites with hole localization either on $Mn3d$ or $O2p$
orbitals. Proceeding from symmetry and energetic consideration we may
conjecture the non-isovalent substitution in doped manganites like
$La_{1-x}Sr_{x}MnO_3$ should result in an inhomogeneous mixture of different
phase states with predominant weight of a phase providing the most efficient
screening of the charge inhomogeneity, in particular, either metallic, or
 metallic-like  phase.
In this connection we consider below a mixed valence phase being the analog of
the well known in solid state chemistry disproportionation reaction, or, in a
sense, electron-hole liquid in semiconductors like $Ge$ and $Si$.

\subsubsection{Electron-hole Bose liquid in manganites} Parent manganites  such
as $LaMnO_3$ are antiferromagnetic insulators with the charge transfer
 gap.  Phase transition to novel hypothetically metallic state  could be realized due to a
  mechanism familiar to such semiconductors with filled bands as Ge and Si where
  given  certain conditions one observes
   a formation of metallic EH-liquid as a result of the exciton decay
   \cite{Rice}. Fundamental absorption band in parent manganites is formed both by the
   intracenter $O2p-Mn3d$ transfer and by the small intercenter charge transfer
    excitons, which  in terms of chemical notions represent somewhat like the disproportionation
     quanta
 with threshold $\sim 3.5$ eV, resulting in a formation of electron $MnO_6^{10-}$ and hole
 $MnO_6^{8-}$ centers.  The excitons may be considered to be well defined entities only at
 small content, whereas at large densities their coupling is screened and their overlap becomes
  so considerable  that they loose individuality, become unstable regarding the decay
  (dissociation) to  electron  and hole  centers, and we come to the system of electrons and
  holes, which form a so-called electron-hole plasma.  Maximally dense EH plasma corresponds
  to overall disproportionation  with transition to novel phase state with mixed valence which
   could be named as EH liquid \cite{Rice}. EH liquid in conventional semiconductors like
   Ge, Si is a two-component {\it Fermi-liquid} whereas EH liquid in manganites would represent
    a system of strongly correlated electron  and hole  centers which is equivalent to a
    {\it Bose-liquid}.

An instability of parent manganite $LaMnO_3$ with regard to overall
disproportionation like (\ref{dispro})
 was strikingly demonstrated recently by Zhou and Goodenough \cite{Zhou}. The transport
 (thermoelectric power and resistivity) and magnetic (susceptibility) measurements showed that
  $LaMnO_3$ above the cooperative Jahn-Teller orbital-ordering temperature $T_{JT}\approx 750$ K
   transforms into charge-disproportionated paramagnetic phase with $\mu _{eff}=5.22 \mu _{B}$
    and cooperative charge transfer of many heavy vibronic charge carriers.

Strong variation of the $LaMnO_3$ Raman spectra with increasing laser power
\cite{Raman} could be related to the photo-induced nucleation and  the volume
expansion of the EH Bose-liquid, especially, as at a rather big excitation
wavelength $\lambda = 514.5$ nm, or $\lambda = 632.8$
 nm the absorption is considerably stronger in domains of novel phase than in
 parent lattice.

Effective nucleation of the EH Bose-liquid in manganites could be provoked by a
non-isovalent substitution since this strongly polarizable or even metallic
phase in contrast with parent insulating phase  provides an effective screening
of charge inhomogeneity.

Even the most relevant quantum-chemical approach to the description of the EH
liquid in
 manganites comes across a number of essential difficulties related to a strong overlap and
 coupling of the neighbouring $MnO_6$ centers, and anomalously strong fluctuations of the
  electron potential to be comparable with the potential itself.

Such an approach starts first of all from effective center with a number of
active valent states
 which form a $valent$ multiplet. Then we introduce a pseudo-spin formalism and an effective
 pseudo-spin Hamiltonian for the system of centers. Various pseudo-spin models are based on
 familiar quantum and classical Heisenberg and Ising models and are widely used for description
  of the charge, spin, orbital and structural ordering in systems with "local" moments. Such an
   approach looks like a conventional variational treatment with trial functions and effective
    parameters which magnitude could be fitted from experiment or numerical quantum-chemical
    $ab-initio$ calculations for rather small model system.

Optimal situation to take account of charge, dipole and quadrupole fluctuations
could be provided if we start from the $MnO_6$ centers, and include to the
valent multiplet of the $effective$ electron and hole $MnO_6$ center the even
orbital singlets ${}^{6}A_{1g}$, ${}^{4}A_{2g}$  and the odd orbital triplets
${}^{6}T_{1u}$, ${}^{4}T_{2u}$, respectively.
 Such effective centers with the ${}^{6}A_{1g}-{}^{6}T_{1u}$, ${}^{4}A_{2g}-{}^{4}T_{2u}$
  valent multiplets could provide optimal simulation for the complex response of the centers
  to different fluctuations of potential. From the other hand, such a model allows to describe
   a rather subtle electron density distribution, and many other static and dynamic properties
    of the EH Bose-liquid. A conventional parametrization approach to appropriate effective
     Hamiltonian does not imply  the detailed information about a microstructure of different
       states involved. However, such an information is essential for quantitative or
       semi-quantitative evaluation of different effective parameters and different terms in
       the Hamiltonian. The relevant choice of electronic configurations for the even
${}^{4}A_{2g}$ and ${}^{6}A_{1g}$ terms of the hole and electron centers in our
case naturally corresponds to $t_{2g}^{3}$ and $t_{2g}^{3}e_{g}^2$ ones,
respectively. For the odd orbital triplets ${}^{4}T_{2u}$, ${}^{6}T_{1u}$ the
most probable candidates are  the low-lying CT configurations
$t_{2g}^{3}e_{g}^{1}\underline{t}_{2u}$ and
$t_{2g}^{4}e_{g}^{2}\underline{t}_{2u}$, respectively.

One should emphasize the specific role of the long-range electrostatic
multipole interactions for the EH Bose-liquid. They can favor stabilization of
an inhomogeneous vortex-like distribution of charge and orbital density.

\subsubsection{"Chemical" approach to an EH Bose-liquid and $Mn$ valence
problem} A simplified "chemical" approach to an EH Bose-liquid as to a
disproportionated phase \cite{Ionov} naively implies an occurrence of
$Mn^{4+}$ and $Mn^{2+}$ ions. However, such an approach is very far from
reality. Indeed, the electron $MnO_6^{10-}$ and  hole $MnO_6^{8-}$ centers are
already the mixed valence centers, as in the former the $Mn$ valence resonates
between $+2$ and $+1$, and in the latter does between $+4$ and $+3$.

In this connection, one should note that in a sense disproportionation reaction
(\ref{dispro}) has several purely ionic counterparts, the two rather simple
$$
Mn^{3+}-O^{2-}-Mn^{3+} \rightarrow Mn^{2+}-O^{2-}-Mn^{4+},
$$
and
$$
Mn^{3+}-O^{2-}-Mn^{3+} \rightarrow Mn^{2+}-O^{1-}-Mn^{3+},
$$
and, finally, one rather complicated
$$
O^{2-}-Mn^{3+}-O^{2-} \rightarrow O^{1-}-Mn^{1+}-O^{1-}.
$$
Thus, the disproportionation (\ref{dispro}) threshold energy has to be
maximally close to the CT energy parameter $\Delta _{pd}$. Moreover, namely
this is seemingly to be one of the main parameters governing
 the nucleation of EH Bose-liquid in oxides.

So far, there has been no systematic exploration of exact valence and spin
state of $Mn$ in
 these systems.
In thermoelectric power experiments (TEP) Hundley and Neumeier \cite{TEP} find
that more hole-like charge carriers or alternatively fewer accessible Mn sites
are present than expected for the value $x$. They suggest a charge
disproportionation model based on the instability of
 $Mn^{3+}-Mn^{3+}$ relative to that of $Mn^{4+}-Mn^{2+}$. This transformation provides excellent
  agreement with doping-dependent trends exhibited by both TEP and resistivity.

Using electron paramagnetic resonance (EPR) measurements Oseroff {\it et al}.
\cite{Oseroff}
 suggest that below 600 K there are no isolated $Mn$ atoms of valence $+2,+3,+4$, however they
 argue that EPR signals are consistent with a complex magnetic entity composed of $Mn^{3+}$ and
  $Mn^{4+}$ ions.

Park et al. \cite{Park} attempted to support the  $Mn^{3+}/Mn^{4+}$ model,
based on the $Mn$ $2p$ x-ray photoelectron spectroscopy (XPES) and $O1s$
absorption. However, the significant discrepancy between the weighted
$Mn^{3+}/Mn^{4+}$ spectrum and the experimental one for given
 $x$ suggests a more complex doping effect. Subias et al.\cite{Subias} examined the valence
 state of $Mn$ utilizing $Mn$ $K$-edge x-ray  absorption near edge spectra (XANES), however,
  a large discrepancy is found between experimental spectra given intermediate doping and
  appropriate superposition of the end members.

The valence state of $Mn$ in Ca-doped $LaMnO_3$ was studied by high-resolution
$Mn$ $K\beta $ emission spectroscopy by Tyson {\it et al.} \cite{Tyson}. No
evidence for $Mn^{2+}$ was observed at any $x$ values seemingly ruling out
proposals regarding  the $Mn^{3+}$ disproportionation. However, this conclusion
seems to be absolutely unreasonable. Indeed, electron center
 $MnO_{6}^{10-}$ can be found in two configuration with formal $Mn$ valence $Mn^{2+}$ and
 $Mn^{1+}$ (not simply $Mn^{2+}$!), respectively. In its turn, the hole center $MnO_{6}^{8-}$
 can be found in two configuration with formal $Mn$ valence $Mn^{4+}$ and $Mn^{3+}$ (not simply
  $Mn^{4+}$ !), respectively. So, within the model the $Mn$ $K\beta $ emission spectrum for
  Ca-doped $LaMnO_3$ has to be a superposition of appropriately weighted $Mn^{1+}, Mn^{2+},
   Mn^{3+}, Mn^{4+}$ contributions (not simply $Mn^{4+}$ and $Mn^{3+}$, as one assumes in
   Ref.2). Unfortunately, we do not know the $Mn$ $K\beta $ emission spectra for the oxide
   compounds  with $Mn^{1+}$ ions, however a close inspection of the $Mn$ $K\beta $ emission
   spectra for the series of  $Mn$-oxide compounds with $Mn$ valence varying from $2+$ to $7+$
   (Fig.2 of the cited paper) allows to uncover a rather clear dependence on valence, and
   indicates a possibility to explain the experimental spectrum for Ca-doped $LaMnO_3$ (Fig.4a)
    as a superposition of appropriately weighted $Mn^{1+}, Mn^{2+}, Mn^{3+}, Mn^{4+}$
    contributions. It should be noted that an "arrested" $Mn$-valence response to the doping
in the $x<0.3$ range founded in Ref. \cite{Tyson} is also consistent with
creation of predominantly oxygen holes.

This set of conflicting data together with a number of additional data
\cite{Croft} suggests the need for an in-depth exploration of the $Mn$ valence
problem  in this perovskite system. However, one might say, the doped
manganites are not only systems with   mixed valence, but systems with
indefinite valence, where we cannot, strictly speaking,  unambiguously
distinguish $Mn$ species with either distinct valence state.

\subsubsection{Effective model Hamiltonian} \label{secham}

The effective model Hamiltonian describing the system of clusters
$MnO_{6}^{8-}$ and $MnO_{6}^{10-}$ with the valent manifolds
$^{4}A_{2g}-{}^{4}T_{2u}$ and $^{6}A_{1g}-{}^{6}T_{1u}$, respectively, can be
presented as a sum of several terms
\begin{equation}
\hat H = \hat H_{QLBG}+\hat H_{orb}+\hat H_{sp}+\hat H_{so}+\hat H_{vibr}.
\label{initham}
\end{equation}
Here $\hat H_{QLBG}$ is the effective Hamiltonian for quantum lattice Bose-gas,
which accounts for quantum transfer of the electron (hole) centers (the
effective kinetic energy of the bosons) and the effective repulsion of these
bosons (see below).

The term $\hat H_{orb}$ is responsible for the crystal field effects,
multipole-multipole interaction, pure orbital part of the exchange interaction
of the centers and the orbital Zeeman energy.
\begin{equation}
\hat H_{orb}=\hat H_{CF}+\hat H_{multi}+\hat H_{orb.ex.}+\hat H_{orb.Zeem.},
\label{hamorb}
\end{equation}
Note that the effective crystal field Hamiltonian $\hat H_{CF}$ includes both
high-symmetry (cubic) term, which is responsible for the
$^{6}A_{1g}-{}^{6}T_{1u}$ ($^{4}A_{2g}-{}^{4}T_{2u}$) splitting, and
low-symmetry term $\hat H_{LSCF}$. Here, one should note the important role of
the long-range multipole-multipole terms in (\ref{hamorb}), leading to
stabilization of the inhomogeneous structures.

The spin Hamiltonian $\hat H_{sp}$ accounts for a spin-spin exchange
interaction and spin Zeeman effect. The term $\hat H_{so}$ accounts for the
spin-orbital interaction. At last, $\hat H_{vibr}$ is responsible for the
vibronic effects, which in presence of the orbital (quasi)degeneracy of the
valent manifold have the (pseudo)Jahn-Teller-like character.

Note that the  oxygen holes  occupying the non-bonding orbitals provide a
strong ferromagnetic exchange with neighboring manganese ions resulting in
formation of strongly coupled $Mn-O-Mn$ ferromagnetic clusters.

The detailed analysis of the effective Hamiltonian (\ref{initham}) represents
itself extremely complex and challenging problem for  further investigation.
Below we shortly discuss one simple version of such a Hamiltonian.

\subsubsection{Simple double-exchange model with triplet bosons}
 In the simplest model of the
manganite as an electron-hole Bose liquid one can restrict himself with orbital
singlets $^{6}A_{1g}$ and $^{4}A_{2g}$ for the electron $MnO_{6}^{10-}$ and
hole $MnO_{6}^{8-}$ centers, respectively. In the absence of the external
magnetic field the effective Hamiltonian of this model takes the form of the
Hamiltonian of the quantum lattice Bose-gas of the triplet bosons (QLTBG)

$$ \hat H = \hat H_{QLBG}+\hat
H_{ex}=-\frac{1}{2}\sum_{<ij>}\tau_{ij}\vec T_{i}^{\dagger}\vec T_{j}
+\sum_{<ij>}V_{ij}N_{i}N_{j}-\mu\sum_{i}N_{i} $$
\begin{equation}
- \frac{1}{2}\sum_{<ij>}J_{ij}^{hh}{\vec S}_{i}{\vec S}_{j} -
\sum_{<ij>}J_{ij}^{hb}{\vec s}_{i}{\vec S}_{j}
-\frac{1}{2}\sum_{<ij>}J_{ij}^{bb}{\vec s}_{i}{\vec
s}_{j}-\sum_{i}J_{ii}^{hb}{\vec s}_{i}{\vec S}_{i}.\label{H}
\end{equation}

Here $T_{\alpha i}^{\dagger}$ denotes the spin $S=1$ and the spin projection
$\alpha$ boson creation operator at the {\it i}-th site; $T_{\alpha i}$ is the
corresponding annihilation operator. The boson number operator
$N_{i}=\vec{T}_{\alpha i}^{\dagger}\vec{T}_{\alpha i}$ at the {\it i}-th site
due to the condition $V_{ii}\rightarrow+\infty$ ({\it hardcore boson}) can take
values 0 or 1. The summation indices $<ij>$ denote sum over the nearest
neighbors.

The operators $T_{\alpha i}^{\dagger}$ and $T_{\alpha i}$ in the $|NSM\rangle$
representation have the following matrix elements \cite{var} :

$$ \left\langle 0\frac{3}{2}m\right|T_{\alpha i}\left|
1\frac{5}{2}m'\right\rangle=(-1)^{\frac{3}{2}-m} \left(
\begin{array}{ccc}\frac{3}{2}&1&\frac{5}{2}
\\
-m&\alpha&m'\end{array}\right)\left\langle\frac{3}{2}\right\|
T\left\|\frac{5}{2}\right\rangle,$$

$$ \left\langle 1\frac{5}{2}m\right|T^{\dagger}_{\alpha i}\left|
0\frac{3}{2}m'\right\rangle=(-1)^{\frac{5}{2}-m} \left(
\begin{array}{ccc}\frac{5}{2}&1&\frac{3}{2}
\\
-m&-\alpha&m'\end{array}\right)\left\langle\frac{5}{2}\right\|
T\left\|\frac{3}{2}\right\rangle.$$

Here $\left\langle\frac{5}{2}\right\|
T\left\|\frac{3}{2}\right\rangle=\left\langle\frac{3}{2}\right\|
T\left\|\frac{5}{2}\right\rangle=\sqrt{6}$ is the reduced matrix element for
these operators. One can deduce that

$$ (\vec{T}^\dagger \vec{T})=N;\quad(\vec{T}
\vec{T}^\dagger)=\frac{3}{2}(1-N) $$

It is noteworthy that the presence of the triplet creation/annihilation
operator with the relations shown above, generally, does not permit to
introduce pseudo-spin formalism conventionally used in the case of QLSBG.

The first term in (\ref{H}) corresponds to the kinetic energy off thee bosons,
$\tau_{ij}$ is the transfer integral. The second one reflects the effective
repulsion ($V_{ij}>0$) of the bosons, located on the neighboring sites.

The chemical potential $\mu$ is introduced to fix the boson concentration
\begin{equation}
n=\frac{1}{N}\sum_{i}\langle N_{i}\rangle
\end{equation}

The last term in (\ref{H}) is the Heisenberg exchange interaction of the spins
of the hole centers with each other (term with $J^{hh}$), spins of the hole
centers with the bosons' spins, located on the nearest sites (term with
$J^{hb}$), and term with $J^{bb}$ accounts for the exchange of the boson spins.

The very last term in (\ref{H}) stands for the intra-center exchange between
the boson spin and the spin of the hole center. In order to account the Hund
rule one should consider it to be infinitely large ferromagnetic. After taking
the limit $J^{hb}_{ii}\rightarrow+\infty$ the spin part oh the (\ref{H}) takes
the form
\begin{equation}
\hat H_{ex}= - \frac{1}{2}\sum_{<ij>}J_{ij}^{hh}{\vec S}^{h}_{i}{\vec
S}^{h}_{j} - \sum_{<ij>}J_{ij}^{he}{\vec S}^{h}_{i}{\vec S}^{e}_{j}
-\frac{1}{2}\sum_{<ij>}J_{ij}^{ee}{\vec S}^{e}_{i}{\vec S}^{e}_{j}. \label{H1}
\end{equation}
Here we introduced the operators of the ''hole center spin'' and ''electron
center spin'' $$ {\vec S}^{h}=(1-N){\vec S};\qquad{\vec S}^{e}=N{\vec S}. $$
New exchange parameters $J^{eh}$ and $J^{ee}$ are related with the old ones by
the linear equations $$ J_{ij}^{eh}=\frac{2}{5}J_{ij}^{hb},\quad
J_{ij}^{ee}=\frac{9}{25}J_{ij}^{hh}+
\frac{6}{25}J_{ij}^{bh}+\frac{4}{25}J_{ij}^{ee}.
$$ At last, let us introduce the hermitian
operators $$ \vec{T}^{x}=\frac{1}{2}(\vec{T}+\vec{T}^{\dagger});\qquad
\vec{T}^{y}=\frac{i}{2}(\vec{T}-\vec{T}^{\dagger})
$$ instead of $T$ and $T^{\dagger}$.

Thus the Hamiltonian will take its final form

$$ \hat{H}=\sum_{<ij>\alpha}\tau_{ij}(\vec{T}_{\alpha
i}^{x}\vec{T}_{\alpha j}^{x}+\vec{T}_{\alpha i}^{y}\vec{T}_{\alpha
j}^{y})+\sum_{<ij>}V_{ij}N_{i}N_{j}-\mu\sum_{i}N_{i} $$
\begin{equation}
- \frac{1}{2}\sum_{<ij>}J_{ij}^{hh}{\vec S}^{h}_{i}{\vec S}^{h}_{j} -
\sum_{<ij>}J_{ij}^{he}{\vec S}^{h}_{i}{\vec S}^{e}_{j}
-\frac{1}{2}\sum_{<ij>}J_{ij}^{ee}{\vec S}^{e}_{i}{\vec S}^{e}_{j}. \label{ham}
\end{equation}

One can easily see that the exchange parameters $J^{hh}$, $J^{eh}$ and $J^{ee}$
play the different role: the exchange integral $J_{eh}$ together with the
parameter $V$ lead to set the different spin multiplicity of the neighboring
sites and thus favor the diagonal long-range order (DLRO) -- charge ordered
phase. The exchange parameters $J_{ee}$ and $J_{hh}$, vice versa, make spin
multiplicity of the neighboring sites to be the same, which establishes droplet
phase.

Generally, this model can be considered as  Bose-analogue of the
double-exchange model \cite{dblex}.

Tentative analysis of  this  model allows to predict  possible phase states
with rather conventional diagonal long-range order (antiferromagnetic
insulating, ferromagnetic metallic), and unconventional nondiagonal long-range
order with superfluidity of triplet bosons and ferromagnetic ordering
\cite{Avvakumov}.


\section{Conclusions}
A problem of adequate electronic model for doped manganites remains
controversial. In such a situation we argue a necessity to discuss all possible
candidate states both for isolated $MnO_6$ centers, and respective systems.
Above we have mainly addressed unconventional charge transfer states for
$MnO_6$ centers with oxygen holes.  The experimental observation  of spectral
changes in optical absorption and  anomalously large effective magnetic moments
in doped manganites both at low ($T\ll T_C$) and high ($T\gg T_C$) temperatures
\cite{Korolyov,Mukhin} provides a clear support for such a "oxygen-hole"
scenario with the well developed fluctuations of the sign of the $O2p-Mn3d$
charge transfer energy. We propose two idealized theoretical models to describe
manganites with unconventional hole behavior. The first implies an overall
oxygen localization for the doped holes. The second assumes formation of the
electron-hole Bose liquid, or a system of the electron $MnO_{6}^{10-}$, and
hole $MnO_{6}^{8-}$ centers. In a sense, this scenario resembles  well-known
disproportionation reaction. In both cases one might expect non-trivial
magnetic behavior with strong ferromagnetic fluctuations due to anomalously
strong ferromagnetic coupling of non-bonding $O2p$ holes with $Mn3d$ electrons.


Electron-hole Bose liquid in doped manganites represents the system of triplet
local bosons moving in the lattice of the hole $S=3/2$ $MnO_{6}^{8-}$ centers
which reveals complex phase diagram both with diagonal and off-diagonal order.

Actual electronic structure in doped manganites is believed to be strongly
inhomogeneous with static and dynamic  fluctuations of different phases
including the above addressed ones. In this connection our consideration allows
to uncover some unusual properties of such fluctuations.

An absence of the adequate physical model for the novel phase to be developed
in  manganites upon the doping  hampers the further progress in complete
understanding and description of the unconventional inhomogeneity in these
systems. Anyway,  making use of either experimental data as a trump in favor of
either theoretical model should
 be made with some caution if  a phase homogeneity of the samples under examination  is
 questionable.

\section{Acknowledgments}

The discussions with N.N. Loshkareva, Yu.P. Sukhorukov, E.A. Ganshina,
 A.V. Korolyov, A.A. Mukhin, P. Novak, D. Khomskii, M. Neumann, V.S. Vikhnin, 
 S.-L. Drechsler, N.G. Bebenin,
  V.R. Galakhov are acknowledged.
The research described in this publication was made possible in part by Award
No.REC-005 of the U.S. Civilian Research \& Development Foundation for the
Independent States of the Former Soviet Union (CRDF). The authors acknowledge a
partial support from the Russian Ministry of Education, grant E00-3.4-280, and
Russian Foundation for Basic Researches, grant 01-02-96404.

\newpage

\begin{center}
{\Large  Figure captions}

\vspace{0.5cm}

Fig.1. Diagram of optically active levels for the $MnO_{6}^{8-}$ center with varying
$\Delta _{pd}$ parameter.
\end{center}


\begin{thebibliography}{99}

\bibitem{DE+JT}
A.J. Millis {\it et al}., Phys. Rev. Lett., {\bf 74} (1995) 5144.

\bibitem{dblex}
P.W. Anderson and H. Hasegawa, Phys. Rev., {\bf 100}, 675 (1955).

\bibitem{dblex1}
P.-G. de Gennes, Phys. Rev., {\bf 118}, 141 (1960).

\bibitem{Millis}
K.H. Ahn, A.J. Millis, cond-mat/9901127 v2

\bibitem{Solovyev}
I.V. Solovyev et al. cond-mat/9903322

\bibitem{Nagaev}
E.L. Nagaev, Physics-Uspekhi {\bf 39}  (1996) 781.

\bibitem{Moskvin}
A.S.Moskvin, Physica B, {\bf 252} (1998) 186.


\bibitem{Khomskii}
D. Khomskii, cond-mat/9909349.

\bibitem{Kugel}
K.I. Kugel, Physics-Uspekhi (to be published).

\bibitem{Jung}
J.H. Jung, K.H. Kim, T.W. Noh et al.,  Phys. Rev., {\bf B57}  (1998) R11043.


\bibitem{Tyson}
T.A. Tyson, Q.Qian, C.-C. Kao et al., Phys. Rev., {\bf B60} (1999) 4665.

\bibitem{Korolyov}
A.V. Korolyov, V.Ye. Arkhipov, V.S. Gaviko {\it et al}., JMMM, {\bf 213} (2000)
63.

\bibitem{Mukhin}
M. Paraskevopoulos, F. Mayr, J. Hemberger {\it et al}., J. Phys.:Condens.
Matter, {\bf 12} (2000) 3993.

\bibitem{Ber}
I. B. Bersuker, V. Z. Polinger, Vibronic interactions in molecules and
crystals. Springer-Verlag, 1989.

\bibitem{Gen}
Gen Matsumoto, J. Phys. Soc. Jap. {\bf 29} (1970) 615.

\bibitem{Zhou}
J.-S. Zhou and J.B. Goodenough, Phys. Rev., {\bf B60} (1999) R15002.

\bibitem{Davydov}
A.S. Davydov, Theory of Molecular Excitons, McGraw-Hill, New York, 1962.

\bibitem{Paulusz}
A.G. Paulusz and H.I. Burrus, Chem. Phys. Lett., {\bf 17} (1972) 527.

\bibitem{MnO}
Ya. M. Ksendzov, M.I. Klinger, I.N. Ivanova {\it et al}., Izv. Akademii nauk SSSR, ser. fiz., {\bf 35}
 (1971) 1178 (in Russian).

\bibitem{Balykina}
E.A. Balykina, E.A. Ganshina, G.S. Krinchik, A.Yu. Trifonov, I.O. Troyanchuk,
JMMM {\bf 117}  (1992) 259.


\bibitem{Zenkov}
A.S. Moskvin, A.V. Zenkov, E.I. Yuryeva, V.A. Gubanov, Physica B, {\bf 168} (1991) 187.

\bibitem{Vikhnin}
V.S. Vikhnin, Ferroelectrics, {\bf 199} (1997) 25.

\bibitem{Moskvin1993}
 A.S. Moskvin, JETP Lett., {\bf 58} (1993) 342.

\bibitem{Moskvin1994}
 A.S. Moskvin, N.N. Loshkareva, Yu.P. Sukhorukov, M.A. Sidorov, A.A. Samokhvalov, JETP,
 {\bf 105}, 967 (1994).


\bibitem{Hirsch}
J.E. Hirsch and S. Tang, Phys.Rev., {\bf B 40}, 2179 (1989);
J.E. Hirsch, in "Polarons and bipolarons in high-$T_c$ superconductors and
related materials",
 eds E.K.H. Salje, A.S. Alexandrov and W.Y. Liang, Cambridge University Press, 1995, p. 234.

\bibitem{Nataf}
R. Nataf, Les modeles en spectroscopie nucleaire, Dunod, Paris, 1965.

\bibitem{Rice}
T.M. Rice, in Solid State Physics, Eds.H. Ehrenreich, F. Seitz, D. Turnbull,
{\bf 32}, 1 (1977).

\bibitem{Raman}
M.N. Iliev, M.V. Abrashev, H.-G. Lee et al.,  Phys. Rev., {\bf B57} (1998)
2872.

\bibitem{Ionov}
S.P. Ionov, G.V. Ionova, V.S. Lubimov, E.F. Makarov,  Phys. Stat. Sol. (b),
{\bf 71} (1975)  11.

\bibitem{TEP}
M.F. Hundley, J.J. Neumeier, Phys. Rev., {\bf B55} (1997) 11511.

\bibitem{Oseroff}
S.B. Oseroff, M. Torikachvili, J. Singley et al., Phys. Rev., {\bf B53} (1996)
6521.

\bibitem{Park}
J.-H. Park, C.T. Chen, S.-W. Cheong et al.,Phys. Rev. Lett., 76 (1999) 4215.

\bibitem{Subias}
G. Subias, J. Garcia, M.G. Proietti et al., Phys. Rev., {\bf B56} (1997) 8183.

\bibitem{Croft}
M.Croft, D. Sills, M. Greenblatt et al., Phys. Rev., {\bf B55} (1997) 8726;
R.S. Liu, J.B. Wu, C.Y. Chang et al., J. Sol. St. Chem., {\bf 125} (1996) 112.

\bibitem{Micnas}
R. Micnas, J. Ranninger, S. Robaszkiewicz, Rev. Mod. Phys {\bf 62}, 113 (1990).

\bibitem{var}
 D. A. Varshalovich, A. N. Moskalev, V. K.
Khersonskii. Quantum Theory of Angular Momentum (World Scientific, Singapore,
1988).
\bibitem{Avvakumov}
I.L. Avvakumov, A.S. Moskvin, unpublished

\end{thebibliography}
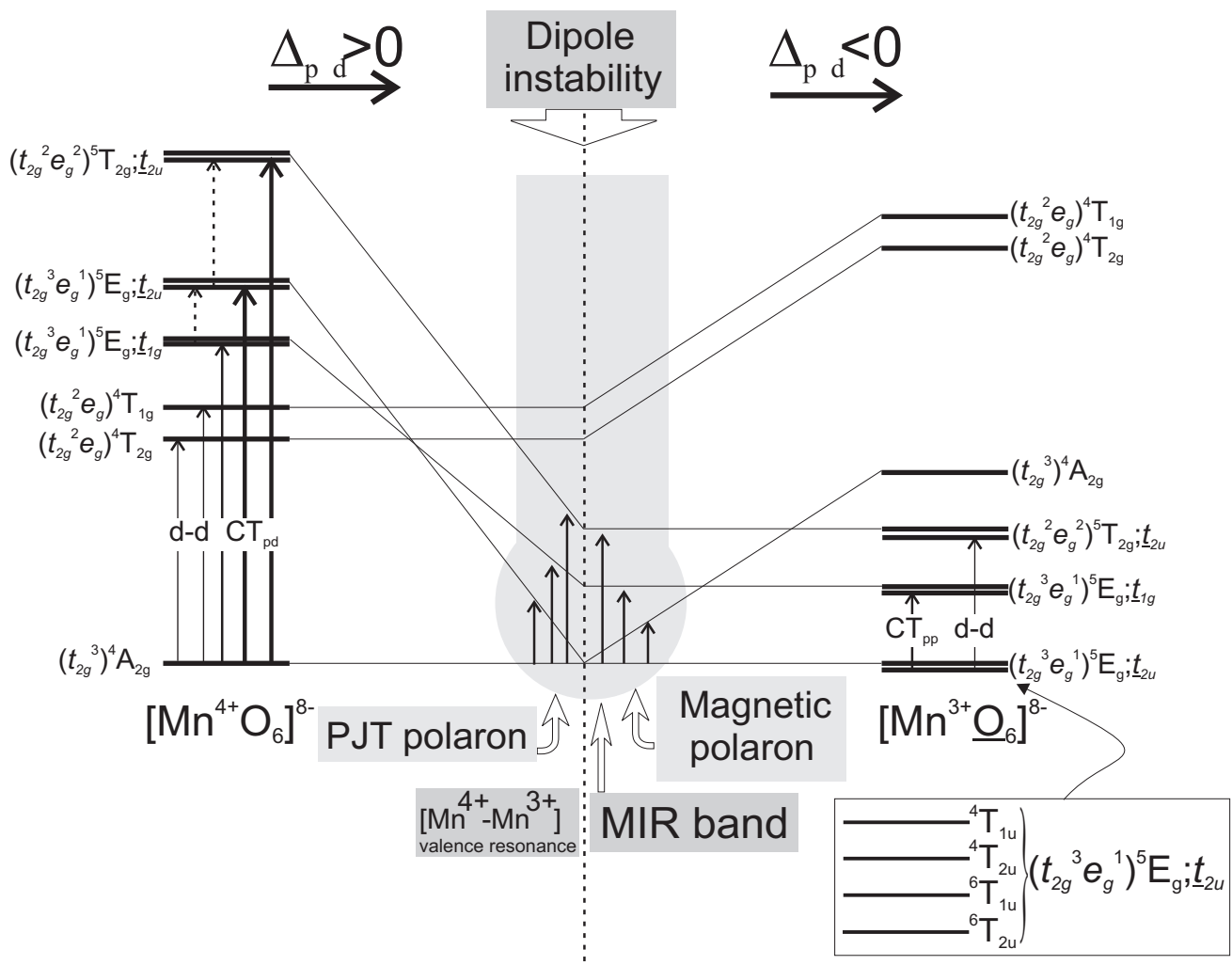
\end{document}